\documentclass[12pt]{iopart}
\usepackage{amsfonts}
\usepackage{amsthm}
\usepackage{amssymb}
\usepackage{graphicx}

\newcommand{\be}{\begin{equation}}

\newcommand{\ee}{\end{equation}}

\def\beqa{\begin{eqnarray}}

\def\eeqa{\end{eqnarray}}

\def\beq{\begin{equation}}

\def\eeq{\end{equation}}

\def\gd{g_{\mu\nu}}

\def\dab{_{\alpha\beta}}

\def\udeab{^{;\alpha\beta}}

\def\pr{{\it Phys. Rev.}\ }

\def\prl{{\it Phys. Rev. Lett.}\ }

\def\pl{{\it Phys. Lett.}\ }

\def\modpl{{\it Mod. Phys. Lett.}\ }

\def\ijmp{{\it Int. Journ. Mod. Phys.}\ }

\def\cqg{{\it Class. Quantum Grav.}\ }

\def\grg{{\it Gen. Relativ. Grav.}\ }

\def\nat{{\it Nature}\ }

\def\apj{{\it Ap. J.}\ }

\def\ncim{{\it Il Nuovo Cim.}\ }

\def\ptp{{\it Prog. Theor. Phys.}\ }

\def\etal{{\it et al.}}

\def\lrr{{\it Liv. Rev. Rel.}\ }

\def\nat{{\it Nature}\ }

\def\apj{{\it Ap. J.}\ }

\let\gam=w

\let\alp=\alpha

\def\gd{g_{\mu\nu}}

\renewcommand{\epsilon}{\varepsilon}

\begin{document}


\title{Cosmological dynamics of $R^n$ gravity}

\author{S Carloni\dag, P K. S. Dunsby\ddag, S Capozziello\S, A Troisi$\|$}

\address{\dag \ddag\ Department of Mathematics and Applied\
Mathematics, University of Cape Town, South Africa.}

\address{\ddag\ South African Astronomical Observatory,
Observatory Cape Town, South Africa.}

\address{\S\ Dipartimento di Scienze Fisiche e Sez. INFN di Napoli,
Universita' di Napoli "Federico II", Complesso Universitario di
Monte S. Angelo, Via Cinthia I-80126 Napoli (Italy)}

 \address{$\|$\ Dipartimento di Fisica``E. R.
Caianiello'', Universit\`{a} di Salerno, I-84081 Baronissi,
Salerno e Istituto Nazionale di Fisica Nucleare, sez. di Napoli,
Gruppo Collegato di Salerno, Via S. Allende-84081 Baronissi (SA),
Italy}

\begin{abstract}

A detailed analysis of dynamics of cosmological models based on
$R^{n}$ gravity is presented. We show that the cosmological
equations can be written as a first order autonomous system and
analyzed using the standard techniques of dynamical system theory.
In absence of perfect fluid matter, we find exact solutions whose
behavior and stability are analyzed in terms of the values of the
parameter $n$. When matter is introduced, the nature of the
(non-minimal) coupling between matter and higher order gravity
induces restrictions on the allowed values of $n$. Selecting such
intervals of values and following the same procedure used in the
vacuum case, we present exact solutions and analyze their
stability for a generic value of the parameter $n$.  From this
analysis emerges the result that for a large set of initial
conditions an accelerated expansion is an attractor for the
evolution of the $R^n$ cosmology. When matter is present a
transient almost-Friedman phase can also be present before the
transition to an accelerated expansion.
\end{abstract}


\maketitle
\section{Introduction}

At the present time, there are many good reasons to consider General
Relativity (GR) as our best theory for the gravitational interaction.
It accounts for the shortcomings of Newtonian gravity and fits
beautifully all the tests that we have been able to devise at laboratory and Solar System levels.
Nevertheless, in the last few decades, there is more and more evidence
that GR may be incomplete. In particular, the difficulties to
match GR with quantum theory, the flattening of galactic
rotation curves and, more recently, the cosmic acceleration can be considered hints that some aspects of
gravitational interaction are still unknown.

It is for these reasons that much recent effort has been spent on the study of
generalizations of the Einstein theory and in particular the so-called Extended Theory of Gravitation (ETG).

One of the most interesting classes of ETG, called {\em non-linear
gravity theories} or {\em higher-order theories of gravity}, is
based on gravitational actions  which are non-linear in the Ricci
curvature ($R$) and$/$or contain terms involving combinations of
derivatives of $R$ \cite{kerner,teyssandier,magnanoff}.  The
higher order theories of gravity arise in a wide range of
different frameworks. For example, it can be shown that for
quantum field theory in curved spacetime, the renormalization of
the stress energy tensor implies the introduction of higher order
corrections to the GR gravitational action \cite{birrell}. In
addition, the low energy limit of $D=10$ superstring theory is a
theory of gravity with additive higher order corrections to the
Hilbert-Einstein (HE) term \cite{stringhe,Ohta}. Finally, the
general form of the vacuum action for Grand Unification Theories
(GUT) is a higher order theory and the GUT models obtained by
relaxing the constraint of linear gravity have, in general, better
physical properties than the standard GR \cite{kalasnikov}.

Unfortunately most of the efforts to achieve an insight into the
physics of these models has been frustrated by the complexity of
the field equations. In fact, the presence of non-linearities and
higher order terms makes it hard to achieve both exact and
numerical solutions which  can be compared with observations. Such
difficulties, added to other problems connected with the presence
of non-gaugeable ghosts in the particle spectrum of these theories
\cite{salam}, have led  researchers to pay more attention towards
other alternative gravity theories such as braneworld models
\cite{maartens}.

The recent realization that the universe is currently undergoing
an accelerated expansion phase and the quest for the nature of the
quintessence field has renewed the interest in higher order
theories of gravity and their relation to cosmology. This is due
to the fact that the higher order corrections to the HE action can
be viewed as an effective fluid and this fluid can emulate the
action of the homogeneous part of the quintessence field
\cite{revnostra}. Hence, in this {\em Curvature Quintessence}
scenario, what we observe as a new component of the cosmic energy
density is an effect of higher order corrections to the HE
Lagrangian. This approach has several advantages over the standard
quintessence scenario. For example, we do not need to search for
a quintessence scalar field and such theories are more easily constrained
by observations.

A key element of the curvature quintessence scenario is the choice
of the form of the Lagrangian density $\mathcal{L}$ for the
gravitational interaction. A simple choice is a generic power law
of the Ricci scalar $R$. This {\em $R^{n}$-gravity} model has been
shown to be a good candidate for describing the present evolution
of the Universe
\cite{revnostra,allemandi,carrol,flanagan,odintsov}. Comparison
with data coming from supernovae data and the age of the universe
as measured by WMAP \cite{wmap} shows that, at least in the vacuum
case, these models are both in agreement with the observations and
allow accelerating solutions \cite{curvatura obs}. This result is
very important, but incomplete. In fact, we do not know what the
stability of these solutions are and which role they play in the
global behavior of the underlying cosmological model. In addition,
when perfect fluid matter is introduced, the equations become even
more complex, making it extremely difficult to find exact
solutions.

These problems can be (at least partially) addressed using the
dynamical system approach. This powerful method was first developed
by Collins and then by Ellis and Wainwright (see \cite{ellisbook}
for a wide class of cosmological models in the GR context). Some
work has also been done in the case of scalar fields in cosmology and for
scalar-tensor theories of gravity \cite{wands,coley,gunzig}. Studying
cosmologies using the dynamical systems approach has the advantage
of offering a relatively simple method to obtain exact solutions
(even if these only represent the asymptotic behavior) and to obtain a
(qualitative) description of the global dynamics of these models.
Such results are very difficult to obtain by other methods.

The purpose of this paper is to apply the dynamical system
approach to higher order gravity and in particular to the
$R^{n}-$gravity models. The paper is organized as follows: In
section \ref{Equazioni e convenzioni}, we present the basic
equations of the model. In section \ref{dynamica1}, we analyze
dynamics of $R^{n}-$gravity in the vacuum case, find exact
solutions and determine their stability. Section \ref{dynamica2}
gives a generalization of these results to the matter case.
Finally, section \ref{conclusioni} contains a summary of the
results and conclusions.

Unless otherwise specified, we will use natural units
($\hbar=c=k_{B}=8\pi G=1$) throughout the paper, Greek indices
run from 0 to 3 and Latin indices run from 1 to 3.
The symbol $\nabla$ represents the usual covariant
derivative and $\partial$ corresponds to partial
differentiation. We use the $(+,-,-,-)$ signature and the
Riemann tensor is defined by
\begin{equation}
R^{\rho}_{\mu\lambda\nu}=W_{\mu\nu,\lambda}^{\rho}-W_{\mu\lambda,\nu}^{\rho}+
W_{\mu\nu}^{\alp}W_{\lambda\alp}^{\rho}+
W_{\mu\lambda}^{\beta}W_{\nu\beta}^{\rho}\;,
\end{equation}
where the $W_{\mu\nu}^{\rho}$ is the Christoffel symbol (i.e. symmetric in the lower indices),
defined by
\begin{equation}
W_{\mu\nu}^{\rho}=\frac{1}{2}g^{\rho\alp}
\left(g_{\mu\alp,\nu}+g_{\alp\nu,\mu}-g_{\mu\nu,\alp}\right)\;.
\end{equation}
The Ricci tensor is obtained by contracting the {\em first}
and the {\em third} indices
\begin{equation}\label{Ricci}
R_{\mu\nu}=g^{\rho\lambda}R_{\rho\mu\lambda\nu}\;.
\end{equation}
\section{Basic Equations} \label{Equazioni e convenzioni}

If we relax the assumption of linearity in the
gravitational action, the general coordinate invariance allows
infinitely many additive terms to the HE action:
\begin{equation}\label{azione generale high order}
S = \int d^4 x \sqrt{-g} \left\{ \Lambda + c_{0} R + c_{1} R^{2} + c_{2}
R_{\mu \nu} R^{\mu \nu} \ldots + {\cal L}_{M} \right\}\;,
\end{equation}
where ${\cal L}_{M}$ is the matter Lagrangian and we have included
corrections up to the fourth order (the fourth order term
$R_{\mu \nu \alpha \beta} R^{\mu \nu \alpha \beta}$ has been neglected
as a consequence of the Gauss-Bonnet theorem.

The action (\ref{azione generale high order}) is not canonical
because the Lagrangian function contains derivatives of the
canonical variables of order higher than one. This means that, not only do
we expect higher order field equations, but also the validity of the
Euler-Lagrange equations is compromised.

This problem is particularly difficult in the general case, but
can be solved for specific metrics. In fact, once the components
of the metric tensor are specified, we can define a new set of
generalized coordinates in such a way to reduce formally the
degree of the Lagrangian. In these new variables, the higher order
field equations correspond to a system of second order
differential equations together with a constraint equation
associated with the definition of the variables themselves
\cite{SalvRev}. As will be clear in the following, this procedure
is crucial in order to both write the field equations for
$R^{n}$-gravity and to apply the dynamical system approach.

In homogeneous and isotropic spacetimes, the Lagrangian in
(\ref{azione generale high order}) can be further simplified.
Specifically, the variation of the term $R_{\mu\nu}R^{\mu\nu}$ can
always be rewritten in terms of the variation of $R^2$ \cite{de
Witt 1965}. Thus, the ``effective" fourth order Lagrangian in
cosmology contains only powers of $R$ and we can suppose, without
loss of generality, that the general form for a non-linear
Lagrangian will be a {\em generic function of  the scalar
curvature}:
\begin{equation}\label{lagr f(R)}
L=\sqrt{-g}\left( f(R)+{\cal L}_{M}\right)\;.
\end{equation}

By varying \footnote{In the present work we will adopt a ``metric
approach to derive the field equations. Alternatively, one can use
the ``Palatini approach" in which the fields ``$g$" and ``$W$" are
considered independent \cite{allemandi,odintsov}.} equation
(\ref{lagr f(R)}), we obtain the fourth order field equations
\beq\label{3.7.2}
f'(R)R_{\mu\nu}-\frac{1}{2}f(R)g_{\mu\nu}=f'(R)\udeab\left(g_{\alpha\mu}g_{\beta\nu}-
g\dab\gd\right)+\tilde{T}^{M}_{\mu\nu}\,, \eeq where
$\displaystyle{\tilde{T}^{M}_{\mu\nu}=\frac{2}{\sqrt{-g}}\frac{\delta
(\sqrt{-g}L_{M})}{\delta g_{\mu\nu}}}$ and the prime denotes a
derivative with respect to $R$. It is easy to check that standard
Einstein equations are immediately recovered if $f(R)=R$. When
$f'(R)\neq0$ the equation (\ref{3.7.2}) can be recast in the more
expressive form
\begin{equation}\label{73-curv3}
G_{\mu\nu}=R_{\mu\nu}-\frac{1}{2}g_{\mu\nu}R=T^{TOT}_{\mu\nu}=T^{R}_{\mu\nu}+T^{M}_{\mu\nu}\,,
\end{equation}
where
\begin{equation}
\label{74-curv4}
T^{R}_{\mu\nu}=\frac{1}{f'(R)}\left\{\frac{1}{2}g_{\mu\nu}\left[f(R)-Rf'(R)\right]+
f'(R)^{;\alpha\beta}(g_{\alpha\mu}g_{\beta\nu}-g_{\alpha\beta}g_{\mu\nu})
\right\}\;,
\end{equation}
and
\begin{equation}
\label{75-curv5}
T^{M}_{\mu\nu}=\frac{1}{f'(R)}\tilde{T}^{M}_{\mu\nu}\,,
\end{equation}
is an effective stress-energy tensor for standard matter. This step is
conceptually very important: we have passed from a model in which
gravity has a complicated structure to a model in which the gravitational field
has the standard GR form with a source made up of two fluids:
perfect fluid matter and an effective fluid ({\em curvature fluid}) that
represents the non-Einsteinian part of the gravitational
interaction \footnote{Obviously we do not need to restrict
ourselves to the $f(R)$ Lagrangians to make this transformation.
In general, we could have started with (\ref{azione generale high
order}) at whatever order and define an effective stress energy
tensor in such a way that the field equations have the form
(\ref{73-curv3}).}.

If we adopt the above picture, there is an important issue related
to the conservation properties of the two new fluids. In fact,
looking at the Bianchi identities for equation (\ref{73-curv3}), we
expect a conservation equation with interaction
terms between the curvature fluid and standard matter. However
this is not the case. In fact, writing down the Bianchi identities we have
\begin{equation}\label{Bianchi}
    0=T_{\mu\nu}^{TOT;\nu}=T_{\mu\nu}^{R;\nu}-\frac{f''(R)}{f'^{2}(R)}R^{,\nu}\tilde{T}_{\mu\nu}^{M}+
\frac{1}{f'(R)}\tilde{T}_{\mu\nu}^{M;\nu}.
\end{equation}
Using the field equations and the definition of the Riemann
tensor, it is easy to show that the sum of the first two terms of
the RHS of the previous expression is zero. Thus,
$T_{\mu\nu}^{TOT;\nu}\propto \tilde{T}_{\mu\nu}^{M}$ and the total
conservation equation reduces to the one for matter only. A
general, independent proof of this result has been given by
Eddington \cite{eddington book}  and then by others (like in
\cite{lovelock book}). They showed that the first variation for
the gravitational action is divergence free regardless of the form
of the invariant that we choose for the Lagrangian. This means
that no matter how complicate the effective stress energy tensor
$T^{TOT}_{\mu\nu}$ is, it will always be divergence free if
$\tilde{T}_{\mu\nu}^{M;\nu}=0$. As a consequence, the total
conservation equation reduces to the one for standard matter only.
This is a very important feature of the curvature fluid and it is
the fundamental reason why we can use the Friedmann metric in
these models even though the curvature fluid may in general
possess heat flux and anisotropic pressure  \cite{Taylor}.

This paper will focus on a particular higher order model: $R^{n}$-gravity,
whose action for the gravitational interaction reads
\begin{equation}\label{71-curv1}
{\cal A}=\int d^4x \sqrt{-g} \left[\chi(n) R^{n}+{L}_{M} \right]\;,
 \end{equation}
where $\chi(n)$ is a function of $n$ that reduces to $1$ for
$n=1$. Whatever the value of $n$ we will suppose that the sign of
$\chi$ is kept positive in such a way that the attractive
character of the gravitational interaction is not compromised.

The field equations for this theory read
\begin{eqnarray}\label{equazioni di campo Rn}
&& \nonumber nR^{n-1}G_{\mu\nu}=\chi(n)^{-1}\tilde{T}_{\mu\nu}^{M}
+\frac{1}{2}g_{\mu\nu}(1-n)R^{n}
+\left[n(n-1)R^{n-2}R^{;\alpha\beta}\right.\\
&&\left.+n(n-1)(n-2)R^{n-3}R^{;\alpha}
R^{;\beta}\right](g_{\alpha\mu}g_{\beta\nu}-g_{\alpha\beta}g_{\mu\nu})
\end{eqnarray}
and for $R\neq0$ they can be recast as
\begin{eqnarray}\label{equazioni di campo Rn Rdiv0}
 \nonumber&&G_{\mu\nu}=  T_{\mu\nu}^{M}+ T_{\mu\nu}^{R}
=\frac{\tilde{T}_{\mu\nu}^{M}}{n\chi(n) R^{n-1}}+\frac{1}{2n}g_{\mu\nu}(1-n)R +\left[(n-1)\frac{R^{;\alpha\beta}}{R}
\right.\\
&&\left.+(n-1)(n-2)\frac{R^{;\alpha}
R^{;\beta}}{R^{2}}\right](g_{\alpha\mu}g_{\beta\nu}
-g_{\alpha\beta}g_{\mu\nu})\;.
\end{eqnarray}
In this way, the non-Einsteinian part of the gravitational
interaction can be modelled as an effective fluid which, in general,
posses thermodynamic properties different from standard matter.

In the Friedmann-Lemaitre-Robertson-Walker (FLRW) metric, the above equations can be written as
\begin{eqnarray}\label{Raychaudhuri Rn}
\nonumber &&2n\frac{\ddot{a}}{a}+n(n-1)H\frac{\dot{R}}{R}
+n(n-1)\frac{\ddot{R}}{R}+n(n-1)(n-2)\frac{\dot{R}^{2}}{R^{2}}-(1-n)\frac{R}{3}\\
&&+\frac{\rho}{3n\chi(n)R^{n-1}}(1+3w)=0\;,
\end{eqnarray}
\begin{equation}\label{friedmann Rn}
H^{2}+\frac{\kappa}{a^{2}}+H\frac{\dot{R}}{R}(n-1)-\frac{R}{6n}(1-n)-\frac{\rho}{3n\chi(n)R^{n-1}}=0\;.
\end{equation}
with
\begin{equation}\label{R FLRW}
    R=-6\left(\frac{\ddot{a}}{a}+\frac{\dot{a}^{2}}{a^{2}}+ \frac{\kappa}{a^{2}}\right)
\end{equation}
where $H=\dot{a}/a$, $\kappa$ is the spatial index and we have
assumed the standard matter to be a perfect fluid with a
barotropic index $w$. Note that in the above equations we have
considered $a$ and $R$ as two independent fields. Such a  position
is standard in canonical quantization of higher order
gravitational theories \cite{FRW nonlin} and is very useful in
this case because it allows to write the  equations
(\ref{equazioni di campo Rn}) as a system of second order
differential equations.

The system (\ref{Raychaudhuri Rn}-\ref{R FLRW}) is completed by
the conservation equation for matter, which as we already pointed
out, is the same as standard GR, i.e.
\begin{equation}\label{conservazione Rn}
\dot{\rho}+3H\rho (1+w)=0 \;.
\end{equation}
We start first with an analysis of the vacuum case and then present the general case in which matter is present.
\section{The Vacuum Case} \label{dynamica1}

Equations (\ref{Raychaudhuri Rn}-\ref{friedmann Rn}) are the
starting point of a dynamical analysis of cosmological models
based on $R^{n}$ gravity. In vacuum ($\rho=0$), their form
suggests the following choice of expansion normalized variables:
\begin{eqnarray}\label{dyn variables lin}
&&\nonumber x=\frac{\dot{R}}{R H}(n-1)\;,\\
&&y=\frac{R}{6 n H^{2}}(1-n)\;,\\
&&\nonumber K=\frac{\kappa}{a^{2}H^{2}}=\frac{\kappa}{\dot{a}^{2}}\;.
\end{eqnarray}
The variable $x$ is associated with the rate of variation of the
Ricci curvature, $y$ represents a measure of the expansion
normalized Ricci curvature and $K$ the spatial curvature of the
Friedmann model. Following \cite{starobinski}, if we consider the
Ricci curvature as a scalar field, we can think of $x$ as the
kinetic term for this field and $y$ as its potential. However,
this analogy does not work completely, because $x$ appears only
linearly in the Friedmann equation.

Defining the new time variable $\mathcal{N}=\ln a$, the cosmological equations are
equivalent to the autonomous system
\begin{eqnarray}\label{eq dyn var lin}
\nonumber x'&=&2+x-x^{2}+\frac{y}{1-n}\left(2+n \,x\right)+(2+x)K\;,\\
y'&=&\frac{x \,y}{n-1}+2y\left(\frac{y \,n}{1-n}+2+K\right)\;,\\
\nonumber K'&=&2K(1+\frac{y \,n}{1-n}+K)\;,
\end{eqnarray}
where the prime represents the derivative with respect to our new ``time" coordinate and
the dynamical variables are constrained by
\begin{equation}\label{constraint delle fasi}
1+K+x-y=0\;.
\end{equation}
It is important to stress that even if the cosmological equations
and the dynamical variables are defined for every value of $n$ the
dynamical equations (\ref{eq dyn var lin}) are only valid for
$n\neq 1$. This is due to the procedure we have used to obtain
this system. However, since the case $n=1$ corresponds to standard
GR, we will limit our discussion to $n\neq 1$.
\subsection{Finite analysis}
The system (\ref{eq dyn var lin}) can be further simplified because the constraint
(\ref{constraint delle fasi}) allows to write the equation for $K$ as a
combination of the other two variables $x$ and $y$. In this way, we can
reduce (\ref{eq dyn var lin}) to the two dimensional system
\begin{eqnarray}\label{eq dyn varridotte}
x'=(2+x)y-2 x-2x^2-\frac{(2+n x)y}{n-1}\;,\\
\nonumber y'=\frac{y}{n-1}\left[(3-2n)x-2y+2(n-1)\right]\;,
\end{eqnarray}
and use (\ref{constraint delle fasi}) to determine the value of
$K$. Note that if $y=0$ the above system implies $y'=0$ and the
$x$ axis is an invariant submanifold for the phase space. This
means that if the initial condition for the cosmological model is
$y\neq0$ a general orbit can approach $y=0$ only asymptotically.
As a consequence, there is no orbit that crosses the $x$ axis and
no global attractor can exist.

Setting $x'=0$ and $y'=0$, we obtain five fixed points
\begin{eqnarray}\label{punti fissi vuoto}
\nonumber \mathcal{A}=\{y\rightarrow 0,x\rightarrow 0\},&&\qquad K=-1\;,
\\ \nonumber \mathcal{B}=\{y\rightarrow 0,x\rightarrow -1\}, &&\qquad K=0\;,
\\ \mathcal{C}=\left\{y\rightarrow \frac{4 n-5}{2 n-1},\;x\rightarrow
\frac{2 (n-2)}{2 n-1}\right\},&&\qquad K=0\;,
\\ \nonumber \mathcal{D}=\{y\rightarrow 2
(n-1)^{2},x\rightarrow -2(n-1)\},&&\qquad K= 2n^2 - 2n -1\;,
\end{eqnarray}
with $\mathcal{A}$ being a double solution. We can distinguish two
classes of fixed point. In the case of the first one
($\mathcal{A},\mathcal{B}$), the coordinates are independent of
the value of $n$ and common to all the $R^{n}$ theories. In the
second one ($\mathcal{C},\mathcal{D}$), the coordinates of the
point are $n$-dependent and their position changes with the
parameter $n$. In particular, the coordinates of the
point $\mathcal{C}$ diverge for $n=1/2$.

Merging occurs for the points $\mathcal{C}$ and $\mathcal{B}$ at
$n=5/4$ and for $\mathcal{C}$ and $\mathcal{D}$ at
$n=\frac{1}{2}(1+\sqrt{3})$ and $n=\frac{1}{2}(1-\sqrt{3})$. The
points $\mathcal{A},\mathcal{D}$ would merge at $n=1$, but we will
not consider this case because our system is not defined for this
value of $n$.

The stability of the fixed points can be obtained by a local
stability analysis \cite{Hartman} and this is summarized in Table
\ref{tavola punti fissi vuoto 2} above.
\begin{table}[t]
\caption{Coordinates of the fixed points, eigenvalues, and solutions for $R^{n}$-gravity in vacuum.}
\centering
\bigskip
\begin{tabular}{cccc}
\br Point& Coordinates (x,y) & Eigenvalues \\ \mr
$\mathcal{A}$ & [0,0] & [-2,2]\\
$\mathcal{B}$  &[-1,0]& $\left[2,\frac{4n-5}{n-1}\right]$ \\
$\mathcal{C}$ &$\left[\frac{2 (n-2)}{2 n-1}\;,\frac{4 n-5}{2 n-1}\right]
$& $\left[\frac{5-4n}{n-1}\;,\frac{4n+2-4n^{2}}{2n^{2}-3n+1}\right]$\\
$\mathcal{D}$ &$\left[2(1-n)\;, 2(n-1)^{2}\right]$&
$\left[n-2+\sqrt{3n(3n-4)}\;, n-2-\sqrt{3n(3n-4)}\right]$
\\\br
\end{tabular}

\begin{tabular}{cc}
\br Point& Solution\\ \mr
$\mathcal{A}$ & $a=a_{0}t$\\
\vspace{1mm}$\mathcal{B}$ & $a= a_{0}(t-t_{0})^{1/2}$ (only for $n=3/2$)\\
\vspace{1mm}$\mathcal{C}$ & $a=a_{0}\;t^{\frac{(1-n)(2n-1)}{n-2}}$\\
\vspace{1mm}$\mathcal{D}$  & $\left\{ \begin{array}{ccc} a=\frac{k
t}{2n^{2}-2n-1}  & \mbox{if}&  k \neq 0\\ a=a_{0}t & \mbox{if}&
k=0 \end{array}\right.$\\\br
\end{tabular}
\label{tavola punti fissi vuoto 1}
\end{table}
\begin{table}
\caption{Stability of the fixed points  for $R^{n}$-gravity in vacuum. We refer to the attractive foci as ``spirals "}
\centering
\bigskip
\begin{tabular}{ccccccccc}
\br Point & $n<\frac{1}{2}(1-\sqrt{3})$ &
$\frac{1}{2}(1-\sqrt{3})<n<0$& $0<n<1/2$& $1/2<n<1$& \\ \mr
 $\mathcal{A}$ & saddle& saddle&saddle&saddle\\
 $\mathcal{B}$  & repulsive& repulsive& repulsive & repulsive \\
 $\mathcal{C}$ & attractive &  saddle &saddle & attractive  \\
 $\mathcal{D}$ &saddle & attractive & spiral&spiral\\
\br\\
\end{tabular}
\begin{tabular}{ccccccccc}
\br Point &
 $1<n<5/4$ & $5/4<n<4/3$& $4/3<n<\frac{1}{2}(1+\sqrt{3})$ & $n>\frac{1}{2}(1+\sqrt{3})$  \\ \mr
 $\mathcal{A}$ &saddle& saddle&saddle&saddle\\
 $\mathcal{B}$  & saddle & repulsive& repulsive& repulsive\\
 $\mathcal{C}$ &repulsive &saddle&saddle& attractive \\
 $\mathcal{D}$ &spiral&spiral& attractive& saddle  \\
\br
\end{tabular}
\label{tavola punti fissi vuoto 2}
\end{table}

In  $\mathcal{A}$ the linearized matrix has a positive and a negative eigenvalue which is independent of
the value of $n$ i.e. it is always a saddle.

The point $\mathcal{B}$ has instead only one fixed (positive) eigenvalue and a second
eigenvalue which depends on $n$. Consequently, it is an unstable node for $n<1$ and $n>5/4$
and a saddle otherwise.

In the case of $\mathcal{C}$,  both the eigenvalues depend on $n$. For
$n<\frac{1}{2}(1-\sqrt{3})$, $1/2<n<1$ and $n>\frac{1}{2}(1+\sqrt{3})$ it
corresponds to stable node and for $1<n<5/4$ is represents an unstable node.
For the other values of $n$, it is a saddle point.

Similarly for $\mathcal{D}$, both the eigenvalues depend on $n$.
For $n<\frac{1}{2}(1-\sqrt{3})$, $n>\frac{1}{2}(1+\sqrt{3})$ it is
a saddle point, for $\frac{1}{2}(1-\sqrt{3})<n<0$ and
$4/3<n<\frac{1}{2}(1-\sqrt{3})$ it is a stable node and for the
other values of $n$ the eigenvalues are complex so $\mathcal{D}$
represents an attractive focus\footnote{The values of $n$ for
which the stability of the fixed points changes are called
bifurcations for the system (\ref{eq dyn varridotte}). In what
follows we will not deal with these specific cases leaving this
discussion for a future work.}.

A pictorial representation of the phase space in the relevant
intervals for $n$ is given in figures (1-8). It is clear that in
this plane the $x$ axis represents an invariant submanifold
because no orbit crosses it.

The coordinates of the fixed points (\ref{punti fissi vuoto}) can
be used to find exact solutions for this cosmological model. In
fact, by substituting the definitions (\ref{dyn variables lin}) in
the field equations we obtain
\begin{equation}\label{H-puntifissi}
    \dot{H}=\left(x_{C}+\frac{y_{C}}{n-1}-1\right)H^{2}\;,
\end{equation}
which links the  coordinates of the fixed point to the form of
$H(t)$. If $n\neq1$ and
 \begin{equation}\label{con}
   (n-1)(x_{C}-1)+y_{C}\neq0
\end{equation}
this equation can be integrated to give
\begin{equation}\label{a-punti fissi}
    a=a_{0}(t-t_{0})^{\alpha}\qquad\mbox{where}\qquad
    \alpha=\left(1-x_{C}-\frac{y_{C}}{n-1}\right)^{-1}\;.
\end{equation}
For the point $\cal A$ we have
\begin{equation}\label{soluzione critico A}
     a=a_{0}(t-t_{0})\;,
\end{equation}
which represent a Milne evolution and is independent of the value
of the parameter $n$. For the point $\cal{B}$ we have
\begin{equation}\label{soluzione critico B}
   a=a_{0}(t-t_{0})^{1/2}\;,
\end{equation}
but the direct substitution in the equation reveals that this
solution is valid only for $n=3/2$. For the point $\cal C$ we have
\begin{equation}\label{soluzione critico C}
a=a_{0}\;(t-t_{0})^{\frac{(1-n)(2n-1)}{n-2}}\;,
\end{equation}
which corresponds to a solution obtained elsewhere \cite{revnostra}
by direct integration of the field equations. The difference is that the
dynamical system approach used here allows us to deduce the stability of this
solution. This issue is particularly important when we use the constraint on $n$
from observations \cite{curvatura obs}. For particular values of this parameter,
$\mathcal{C}$ is not only an attractor but also the {\em only} finite attractor
for the phase space.
For the point $\mathcal{D}$ we have again a Milne evolution:
\begin{equation}\label{critico5}
a=a_{0}\;(t-t_{0})\;,
\end{equation}
but this time the value of $n$ influences the behavior of the solution.
In fact for $n\neq\frac{1\pm\sqrt{3}}{2}$ the constant $a_{0}$ is
given by
\begin{equation}\label{a05}
a_{0}^{2}=\frac{k}{2n^{2}-2n-1}\;,
\end{equation}
otherwise the constant $a_{0}$ remains undetermined. It follows
that the value of $n$ determines the evolution rate unless
$n=\frac{1\pm\sqrt{3}}{2}$.

It might seem surprising that these solutions posses {\em less}
than the four integration constants which we would expect when
solving a fourth order differential equation. This fact can be
explained by noting that the solutions for the fixed points are
asymptotic in nature so we can always imagine that the other
constants occur in a transient part of the solution that is not
relevant in the asymptotic regime.

It is also useful to define the deceleration parameter $q$ in
terms of the dynamical variables. We have
\begin{equation}\label{deceleration par}
    q=-\frac{H'}{H^{2}}-1=-x-\frac{y}{n-1}
\end{equation}
which means, for $q>0$,
\begin{equation}
    \left\{%
\begin{array}{ll}
    n>1 \\
    x(n-1)+y<0\\
\end{array}%
\right.
\left\{%
\begin{array}{ll}
    n<1 \\
    x(n-1)+y>0\\
\end{array}%
\right.
\end{equation}
Thus the phase space is divided into two regions which correspond,
once $n$ is fixed, to accelerating and decelerating evolution. It
is clear that these conditions are never satisfied for the points
${\mathcal A},{\mathcal B},{\mathcal D}$. For the point ${\mathcal
C}$ the constraint on $n$ are
\begin{equation}\label{acc c cond}
\left\{%
\begin{array}{cc}
    n>\frac{1}{2}(1+\sqrt{3})\qquad \mbox{expansion}, \\
    \frac{1}{2}<n<1 \qquad \mbox{     expansion},\\
    n<\frac{1}{2}(1-\sqrt{3})\qquad \mbox{contraction}. \\
\end{array}%
\right.
\end{equation}

\subsection{Asymptotic analysis}
Since the phase plane is not compact it is possible that the
dynamical system (\ref{eq dyn varridotte}) has a nontrivial
asymptotic structure. Thus the above discussion would be
incomplete without checking the existence of fixed points  at
infinity (i.e. when the variables $x$ or $y$ diverge) and
calculating their stability. Physically, such points represents
regimes in which one or more of the terms in the Friedmann
equation (\ref{friedmann Rn}) become dominant. For example, if
$x\rightarrow \infty$, the "kinetic" part of the curvature energy
density dominates over the "potential" part. The asymptotic
analysis can easily be performed {\it compactifying} the phase
space using the so called Poincar$\acute{e}$ method. In the vacuum
case, the compactification can be achieved by transforming to
polar coordinates $(r(\cal{N}), \theta (\cal{N}))$:
\begin{equation}\label{coordinate compattificazione}
x\rightarrow r \cos(\theta)\;,\qquad \qquad y\rightarrow r \sin(\theta)\;,
\end{equation}
and substituting $r\rightarrow\frac{{\cal R}}{1-{\cal R}}$ so that
the regime $r\rightarrow\infty$ corresponds to ${\cal
R}\rightarrow 1$. Using the coordinates (\ref{coordinate
compattificazione}) and taking the limit ${\cal R}\rightarrow 1$,
the equations (\ref{eq dyn varridotte}) can be written as
\begin{eqnarray}
\label{rho'}{\cal R}' &\rightarrow& \frac{(9 - 8n)\cos(\theta) -
\cos(3\theta) - 7\sin(\theta) + \sin(3\theta)}{4(n-1)} \\
\label{theta'} {\cal R}\;\theta' &\rightarrow&
\frac{\cos(\theta)\sin(\theta)[\sin(\theta)-\cos(\theta)]}{(n-1)(1-{\cal
R})}\;.
 \end{eqnarray}
Starting from (\ref{rho'}-\ref{theta'}) we can analyze the
asymptotic features of the system (\ref{eq dyn varridotte}) in the
same way as we did in the finite case. Indeed, since equation
(\ref{rho'}) does not depend on the coordinate ${\cal R}$ we can
find the fixed points using only equation (\ref{theta'}). Setting
$\theta'=0$, we obtain six fixed points which are given in Table
(\ref{tavola punti fissi asintotici vuoto 1}). Their angular
coordinate does not depend on the value of the parameter $n$, so
they are the same irrespective of the power of $R$ appearing in
the initial Lagrangian.

The solutions associated with these fixed points are summarized in
Table \ref{tavola punti fissi asintotici vuoto 1}. The points
$\mathcal{A}_{\infty},\mathcal{D}_{\infty}$ and
$\mathcal{C}_{\infty},\mathcal{F}_{\infty}$ are characterized
respectively by the kinetic part or the potential part of the
energy density of the curvature fluid dominating the dynamics. Let
us derive the solution for $\mathcal{A}_{\infty}$. This point
corresponds to $x\rightarrow\pm\infty$, $y\rightarrow 0$. In the
limit $x\rightarrow\infty$ the first equation of the system
(\ref{eq dyn varridotte}) and (\ref{H-puntifissi}) reduce to
\begin{equation}\label{Ainf}
   x'\rightarrow -2x^{2}
\end{equation}
and
\begin{equation}\label{H-puntifissi-Ainf}
    \frac{H'}{H}=x\;.
\end{equation}
Integrating (\ref{Ainf}) and substituting in
(\ref{H-puntifissi-Ainf}) we obtain
\begin{equation}\label{sol Ainf}
|{\cal{N}}-{\cal
N}_{\infty}|=\left[C_{1}\pm\frac{C_{0}}{2}(t-t_{0})\right]^{2}\;,
\end{equation}
which represents a scale factor that approaches a constant when
$t\rightarrow t_{0}$ (${\cal{N}}\rightarrow{\cal N}_{\infty}$).
The same procedure can be applied to the points
$\mathcal{D}_{\infty}, \mathcal{C}_{\infty}, \mathcal{F}_{\infty}$
obtaining the same result.

The points $\mathcal{C}_{\infty}$ and $\mathcal{F}_{\infty}$
represent a state in which the total energy density of the
curvature fluid dominates and corresponds to closed Einstein
universes. For these points $x$ and $y$ goes to infinity at the
same rate ($\theta=\pi/4$). In this limit, the system (\ref{eq dyn
var lin}) reduces to the equation
\begin{equation}\label{eq asy vacuum}
x'=\frac{1-2n}{n-1}x^{2}
\end{equation}
which admits the solution
\begin{equation}\label{x asy vacuum}
    x=\frac{n-1}{(2n-1)({\cal{N}}-{\cal N}_{\infty})}\;,
\end{equation}
where $\cal{N}_{0}$ is an integration constant. In the same way,
the (\ref{H-puntifissi}) can be written as
\begin{equation}\label{H-puntifissi-asy}
    \frac{H'}{H}=\frac{n x}{n-1}\;.
\end{equation}
Substituting (\ref{x asy vacuum}) in (\ref{H-puntifissi-asy}) and integrating we obtain
\begin{equation}\label{H asy vacuum}
    |H|=|\dot{{\cal
    N}}|=C_{0}\left|{\cal{N}}-{\cal N}_{\infty}\right|^{\frac{n}{2n-1}}\;,
\end{equation}
where $C_{0}$ is a positive integration constant. Integrating
again with respect to the variable $t$ we obtain
\begin{equation}\label{sol asy vacuum}
 |{\cal N}-{\cal N}_{\infty}|= \left[C_{1}\pm
 C_{0}\left|\frac{n-1}{2n-1}\right|(t-t_{0})\right]^{\frac{2n-1}{n-1}}/;.
\end{equation}
This solution represents an expansion if $n<1/2$ and $n>1$, a
contraction if $1/2<n<1$.

The value of $n$ influences also the stability of these fixed
points (see Table \ref{tavola punti fissi asintotici vuoto 2}). In
particular, we see that only the points
$\mathcal{B}_{\infty},\mathcal{D}_{\infty},\mathcal{F}_{\infty}$
can be attractors. For the values of $n$ suggested by the
observations only the points $\mathcal{D}_{\infty}$ or
$\mathcal{F}_{\infty}$ are attractors and we can conclude by
saying that, for these values of $n$, the evolution is always
given by (\ref{sol asy vacuum}).
\begin{table}
\caption{Asymptotic fixed points, $\theta$ coordinates and solutions for $R^{n}$-gravity in vacuum.}
\centering
\bigskip
\begin{tabular}{cccl}
\br Point&  $\theta$ & Behavior \\ \mr
$\mathcal{A}_{\infty}$ & 0 & $|{\cal{N}}-{\cal N}_{\infty}|=\left[C_{1}\pm\frac{C_{0}}{2}(t-t_{0})\right]^{2}$\\
$\mathcal{B}_{\infty}$  &$\pi/4$ & $ |{\cal N}-{\cal N}_{\infty}|= \left[C_{1}\pm C_{0}\left|\frac{n-1}{2n-1}\right|
(t-t_{0})\right]^{\frac{2n-1}{n-1}}$\\
$\mathcal{C}_{\infty}$ &$\pi/2$ & $|{\cal{N}}-{\cal N}_{\infty}|=\left[C_{1}\pm\frac{C_{0}}{2}(t-t_{0})\right]^{2}$\\
$\mathcal{D}_{\infty}$ & $\pi$ & $|{\cal{N}}-{\cal N}_{\infty}|=\left[C_{1}\pm\frac{C_{0}}{2}(t-t_{0})\right]^{2}$\\
$\mathcal{E}_{\infty}$  &$5\pi/4$& $|{\cal N}-{\cal N}_{\infty}|= \left[C_{1}\pm C_{0}\left|\frac{n-1}{2n-1}\right|
(t-t_{0})\right]^{\frac{2n-1}{n-1}}$\\
$\mathcal{F}_{\infty}$ &$3\pi/2$ & $|{\cal{N}}-{\cal N}_{\infty}|=\left[C_{1}\pm\frac{C_{0}}{2}(t-t_{0})\right]^{2}$ \\\br\\
\end{tabular}
\label{tavola punti fissi asintotici vuoto 1}
\end{table}
\begin{table}
\caption{Stability of the asymptotic fixed points  for $R^{n}$-gravity in vacuum.}
\centering
\bigskip
\begin{tabular}{cccc}
\br Point & $n<1/2$ & $1/2<n<1$ & $n>1$  \\ \mr
$\mathcal{A}_{\infty}$ & repulsive& repulsive& repulsive\\
$\mathcal{B}_{\infty}$  &  saddle & attractive &saddle \\
$\mathcal{C}_{\infty}$ & saddle& saddle& repulsive \\
$\mathcal{D}_{\infty}$ & attractive& attractive& attractive\\
$\mathcal{E}_{\infty}$ & saddle& repulsive& saddle\\
$\mathcal{F}_{\infty}$  &saddle &  saddle & attractive\\
\br\\
\end{tabular}
\label{tavola punti fissi asintotici vuoto 2}
\end{table}

\vspace{0.5cm} Summarizing the above results, we can say that the
global behavior of vacuum $R^{n}$ gravity exhibits transient Milne
and power law states that eventually approach the solutions
$\mathcal{C}$, $\mathcal{D}$ or
$\mathcal{B}_{\infty},\mathcal{D}_{\infty},\mathcal{F}_{\infty}$
depending on the value of the parameter $n$ and the initial
conditions. For $n$ far from one, only the points $\mathcal{C}$
and $\mathcal{D}_{\infty},\mathcal{F}_{\infty}$ are attractors,
the first one representing an accelerating solution. In the
neighborhood of $n=1$ the physics becomes more complex because
$\mathcal{D}$ and $\mathcal{B}_{\infty}$ can acquire a stable
character (node or focus) and $\mathcal{C}$ changes its stability.
Finally, the phase space analysis shows that, for these type of
theories, the sign of the curvature remains fixed once it has been
set by the initial conditions. This is an important aspect of
dynamics of the vacuum model because, if the coordinate $y$ of the
point $\mathcal{C}$ is different from the sign of the initial
condition for the curvature, the solution $\mathcal{C}$ is never
reached. This means that $\mathcal{C}$ cannot be considered a
global attractor for the vacuum case and an accelerating expansion
phase is not guaranteed.

\section{ The Matter Case} \label{dynamica2}
When matter is present (as a perfect fluid), each of the
cosmological equations gain a new source term containing the
energy density. This  modification has deep consequences because,
as we can see from (\ref{Raychaudhuri Rn}-\ref{friedmann Rn}), the
matter terms are coupled with a generic power of the curvature.
Now, since in general the sign of the Ricci scalar is not fixed,
these terms will not be defined for every real value of $n$. In
particular we can allow only values of $n$ belonging to the set of
the relative numbers $\mathcal{Z}$ and the subset of the rational
numbers ${\mathcal{Q}}$ which can be expressed as fractions with
an odd denominator

In this sense, the presence of matter induces a natural constraint
through the field equations on $R^{n}$-gravity. In other words, if
we suppose the field equations (\ref{73-curv3}) to be valid, not
all the higher order theories of gravity are consistent in
 presence of standard matter. This difference between vacuum
and non-vacuum physics is not present in GR but it is a common
feature of higher order theories.

The constraints on $n$ also introduces technical problems because
some of the results that we will present below require numerical
calculations. This means that the result should be expressed in
terms of the allowed set of values of $n$, but this is neither an
easy nor a useful task. For this reason, we will
work as if $n$ was unconstrained, supposing that all the intervals
that we devise are meant to represent the subset of allowed
values of $n$ within these intervals, and that their boundaries
approximate the nearest allowed value of n.

With this in mind, let us now derive the dynamics for the matter case.
We again start with (\ref{friedmann Rn}), and define the expansion normalized
variables
\begin{eqnarray}\label{dyn variables lin matt}
&&\nonumber x=\frac{\dot{R}}{R H}(n-1)\;,\\
&&y=\frac{R}{6nH^{2}}(1-n)\;,\\
&&\nonumber z=\frac{\rho}{3n \chi(n) H^{2} R^{n-1}}\;,\\
&&\nonumber K=\frac{\kappa}{a^{2}H^{2}}\;,
\end{eqnarray}
where $z$ represents the contribution of standard matter in the
form of the energy density weighted by a power of the Ricci
curvature. In order to write down a dynamical equation for this
variable, we need an expression for the time derivative of the
energy density. Such an expression can be obtained from the
Bianchi Identities $T^{TOT\; ;\nu}_{\;\mu\nu}=0$ which describe
the conservation properties. As we have seen, if
the gravitational field equations (\ref{73-curv3}) hold, it is
possible to show \cite{lovelock book,eddington book}
that  by virtue of the Bianchi identities, whatever the form of the
gravitational Lagrangian, the conservation equation for matter are
simply given by (\ref{conservazione Rn}).

Differentiating (\ref{dyn variables lin matt}) with respect to the
logarithmic time, we obtain the autonomous system
\begin{eqnarray}\label{sistema  materia}
 x'= x-2x^2+\frac{(2+n x)y}{1-n}-(2+x)K+2-z(1+3w)\;,\\
 y'=\frac{x \,y}{n-1}+2y\left(\frac{y \,n}{1-n}+2+K\right)\;,\\
 z'=-3z(1+w)+2z\left(2+\frac{y n}{n-1}+K\right)-x z\;,\\
 K'=2K\left(1+\frac{y n}{n-1}+K\right)\;,
\end{eqnarray}
with the constraint
\begin{equation}\label{constraint materia}
1+x-y-z+K=0\;.
\end{equation}
As before, we can reduce the number of variables by
substituting (\ref{constraint materia}) into the other
equations and recognizing that the spatial curvature equation
is a linear combination of the other three. In this way, the final
system becomes
\begin{eqnarray}\label{sistema compatto materia}
\nonumber x'=\frac{2(n-2)y}{n-1}-2 x-2x^2-\frac{x y}{n-1}+(1+x)z-3zw\;,\\
y'=\frac{y}{n-1}\left[(3-2n)x-2y+2(n-1)z+2(n-1)\right]\;,\\
\nonumber z'=z\left(2z-1-3x-\frac{2y}{n-1}-3w\right).
\end{eqnarray}
Note that, as in the previous system, $y=0\Rightarrow y'=0$, but
now also $z=0\Rightarrow z'=0$ holds. Consequently the two planes
$y=0$ and $z=0$ correspond to two invariant submanifolds. Since
for $z=0$ (\ref{sistema compatto materia}) reduces to (\ref{eq dyn
varridotte}), we would  be tempted to interpret the plane $z=0$ to
be the vacuum invariant submanifold of the phase space. This is
not entirely correct. In order to understand this point, let us
write the energy density  $\rho$ in terms of the variables
(\ref{dyn variables lin matt}). We have
\begin{equation}\label{rho-dyn var}
    \rho=z y^{n-1}H^{2n}\;.
\end{equation}
Form this formula it is clear that when $z=0$ (and $y\neq 0$) the
density is zero whatever the value of the parameter $n$. Instead
if $y=0$ (and $z\neq 0$) the behavior depends on the value of $n$.
In particular, the density is zero for $y=0$ and $n>1$ and
divergent for $n<1$. If {\em both} $x$ and $z$ are equal to zero
for $n<1$ the behavior of $\rho$ is undetermined and can be found
only by direct substitution in the cosmological equations.

As a consequence, for $n>1$ both $(x,y)$ and the $(x,z)$ planes
are invariant vacuum manifolds, but if $n<1$ the vacuum
submanifold is not necessarily compact because it is made by
$(x,y)$ and other (in principle) submanifolds of the $(x,z)$
plane.

The fact that $z=0$ and $y=0$ are invariant submanifolds also
suggests that there are no orbits which can cross these planes: if
we choose initial conditions $y_{i}=0$ ($z_{i}=0$), the equations
(\ref{sistema compatto materia}) imply that the orbits will never
leave the plane $y=0$ ($z=0$). As a consequence, if the evolution
has initial conditions for which $y_{i}\neq 0$ ($z_{i}\neq 0$)
they can approach the plane $y=0$ only asymptotically.
Consequently, like in the vacuum case, no global attractor exists
in these models because no orbit can, in general, cover all of the
phase space.
\subsection{Finite Analysis}
Setting $x'=0$, $y'=0$, $z'=0$ and solving for $x,y,z$ we obtain the fixed points:
\begin{eqnarray}\label{punti fissi materia}
\fl\nonumber \mathcal{A}=\{y\rightarrow 0,\; x\rightarrow 0,\;
z\rightarrow 0\},\;\qquad &&K=-1 ,\; \\ \fl\nonumber
\mathcal{B}=\{y\rightarrow 0,\; x\rightarrow -1,\; z\rightarrow
0\},\; \qquad&& K=0 ,\;
\\ \fl \mathcal{C}=\left\{y\rightarrow \frac{4 n-5}{2 n-1},\;\;x\rightarrow
\frac{2 (n-2)}{2 n-1},\;z\rightarrow 0\right\},\;\qquad &&K=0,\;
\\\fl \nonumber\mathcal{D}=\{y\rightarrow 2 (n-1)^{2},\; x\rightarrow -2(n-1),\;
z\rightarrow 0\},\;\qquad &&K= 2n^2- 2n -1  ,\;
\\ \fl \nonumber \mathcal{E}=\{y\rightarrow 0,\;\;x\rightarrow -1-3w,\;z\rightarrow -1-3w\},\;\qquad && K=-1 ,\;
\\ \fl\nonumber \mathcal{F}=\{y\rightarrow 0,\;\;x\rightarrow 1-3w,\;z\rightarrow 2-3w\},\;\qquad &&K=0,\;
\\\fl\nonumber \mathcal{G}=\left\{y\rightarrow \frac{(n-1)(4n-12w-3)}{2 n^{2}},\;\;x\rightarrow -\frac{3(n-1)(1+w)}{n},\right.\;
\\\nonumber \left.z\rightarrow \frac{
n(13+9w)-2n^{2}(4+3w)-3(1+w)}{2n^{2}}\right\},\;\qquad &&K=0\;.
\end{eqnarray}

Note that the fixed points $\mathcal{A}, \mathcal{B}, \mathcal{C}, \mathcal{D}$ have the same
coordinates (\ref{punti fissi vuoto}) found in the vacuum case,
with the additional condition that the variable $z$ has to be zero.

Also in this case we have two different classes of fixed points, one independent of the
value of the parameters $n$ ($\mathcal{A},\mathcal{B},\mathcal{E},\mathcal{F}$) and one
that changes its coordinates with $n$ ($\mathcal{C}, \mathcal{D}, \mathcal{G}$).

There are also values of $n$ for which  two of the fixed points ($\mathcal{C}, \mathcal{G}$)
acquire an asymptotic character, namely $0,1/2$. However, $n=1/2$ is not an allowed value in
this case and $n=0$ corresponds to a constant Lagrangian for the gravitational interaction.
These values of $n$ are not interesting for our purposes and are therefore excluded.

Merging occurs for $w$ in the Zel'dovic interval, excluding $n=1$, for $\mathcal{A}$
and $\mathcal{B}$ at $n=5/4$, for $\mathcal{C}$ and $ \mathcal{D}$ at
$n=\frac{1}{2}(1+\sqrt{3})$ and $n=\frac{1}{2}(1-\sqrt{3})$, for $\mathcal{C}$
and $\mathcal{G}$ when $n$ is solution of the equation $(8+6w)n^{2}-(13+9w)n+3(1+w)=0$.
None of this values are allowed in the presence of matter, so we can consider these fixed
points to be always distinct.

The stability of the fixed points can be obtained, as usual, by
linearizing the system (\ref{sistema compatto materia}) and are
summarized in Tables (\ref{tavola punti fissi materia 3a},
\ref{tavola punti fissi materia 3c} and \ref{tavola punti fissi
materia 4}). Because the phase space is three dimensional, the
general dynamical behavior is not easy to visualize. We therefore
do not give any graphical representation as in the vacuum case and
focus only on the stability of the fixed points.
\begin{table}
\caption{Coordinates of the fixed points and solutions for $R^{n}$-gravity with matter.}
\centering
\bigskip
\begin{tabular}{cccc}
\br Point& Coordinates $(x,y,z)$ & Scale Factor  \\ \mr
$\mathcal{A}$ & $[0,0,0]$ & $a=a_{0}(t-t_{0})$ & \\
$\mathcal{B}$ & $[-1,0,0]$ & $a=a_{0}(t-t_{0})^{1/2}$ (only for $n=3/2$)\\
$\mathcal{C}$ & $\left[\frac{2 (n-2)}{2 n-1},\frac{4 n-5}{2 n-1},0\right]$ & $a=a_{0}\;t^{\frac{(1-n)(2n-1)}{n-2}}$ \\
$\mathcal{D}$ & $[2(1-n), 2(n-1)^{2}, 0]$ & $\left\{ \begin{array}{ccc} a=\frac{k t}{2n^{2}-2n-1}
& \mbox{if}&  k \neq 0\\ a=a_{0}t & \mbox{if}& k=0 \end{array}\right.$ \\
$\mathcal{E}$ &$[ -1-3w,0,-1-3w]$ & $a=a_{0}(t-t_{0})$\\
$\mathcal{F}$ &$[1-3w,0,2-3w]$ & $a=a_{0}(t-t_{0})^{1/2}$ (only for $n=3/2$)\\
$\mathcal{G}$ &
$\left[-\frac{3(n-1)(1+w)}{n},\frac{(n-1)[4n-3(w+3)]}{2 n^{2}},
\right.$ & \\  &$\left.\frac{
n(13+9w)-2n^{2}(4+3w)-3(1+w)}{2n^{2}}\right]$ &$a=a_{0}\;
t^{\frac{2n}{3(1+w)}}$\\  \br \\
\end{tabular}
\begin{tabular}{cc}\br Point & Matter Density \\\mr
$\mathcal{A}$  & $\rho=0$\\
$\mathcal{B}$   & $\rho=0$\\
 $\mathcal{C}$  & $\rho=0$\\
 $\mathcal{D}$  & $\rho=0$\\
 $\mathcal{E}$  & $\rho=0$ for $n>1$ or $\rho$ divergent for $n<1$ \\
 $\mathcal{F}$  & $\rho=0$\\
$\mathcal{G}$  & $\rho=\rho_{0} t^{-2n},$ \\  &
$\rho_{0}=(-1)^{n}3^{-n}2^{2n-1} n^{n} (1+w)^{-2n}\times$\\
&$(4n-3(1+w))^{n-1} [2n^{2}(4+3w)-n(13+9w)+3(1+w)]$\\\br
\end{tabular}
\label{tavola punti fissi materia 1}
\end{table}

\begin{table}
\caption{Eigenvalues of the fixed points  for $R^{n}$-gravity with matter.}
\centering
\bigskip
\begin{tabular}{cc}
\br Point & Eigenvalues \\ \mr
 $\mathcal{A}$ &  $(-2,2,-1-3w)$ \\
 $\mathcal{B}$  &$\left(2,\frac{4n-5}{n-1},2-3w\right)$ \\
 $\mathcal{C}$ & $\left(\frac{4 n-5}{n-1}, \frac{2+4n-4n^{2} }{1-3n+2n^{2}},
\frac{-3-13n-8n^{2}-3w+9nw-6n^{2}w}{(n-1)(2n-1)}\right)$\\
 $\mathcal{D}$ & $\left(-2+n-\sqrt{3n(3n-4)},-2+n+\sqrt{3n(3n-4)},2n-3(1+w)\right)$ \\
 $\mathcal{E}$ & $(-2,\frac{2n-3-3w}{n-1},1+3w)$\\
 $\mathcal{F}$ & $(2,\frac{4n-3-3w}{n-1},3w-2)$\\
 $\mathcal{G}$ & $\left(\frac{3-2n+3w}{n},\frac{P_{1}(n,w)-
\sqrt{P_{2}(n,w)}}{4n(n-1)},\frac{P_{1}(n,w)+
\sqrt{P_{2}(n,w)}}{4n(n-1)}\right)$,\\  &
$P_{1}(n,w)=\left(3(1+w)+3n((2n-3)w-1)\right)$\\ &
$P_{2}(n,w)=(n-1)[4n^{3}(8+3w)^{2}-4n^{2}(152+3w(55+18w))$\\&
$+3n(1+w)(139+87w)- 81(1+w)^{2}]$\\\br
\end{tabular}
\label{tavola punti fissi materia 2}
\end{table}

\begin{table}
\caption{Stability of the fixed points for $R^{n}$-gravity with matter (dust or radiation). The term
``spiral" has been used for  pure attractive focus-nodes, the term ``anti-spiral" for pure repulsive
focus-nodes, and  the term``saddle-focus" for unstable focus-nodes. The point $\mathcal{G}$ has been
treated separately because of the approximations used.}
\centering
\bigskip
\begin{tabular}{ccccc}
\br & $n<\frac{1}{2}(1-\sqrt{3})$ & $\frac{1}{2}(1-\sqrt{3})<n<0$
& $0<n<1/2$& $1/2<n<1$
     \\ \mr
 $\mathcal{A}$ & saddle& saddle& saddle& saddle\\
 $\mathcal{B}$  & repellor& repellor& repellor& repellor \\
 $\mathcal{C}$ & attractor& saddle& saddle& attractor \\
 $\mathcal{D}$ & saddle& attractor &  spiral &  spiral\\
 $\mathcal{E}$ & saddle& saddle& saddle& saddle\\
 $\mathcal{F}$ & saddle& saddle& saddle& saddle\\
\br \\
\end{tabular}
\begin{tabular}{ccccc}
  \br  &$1<n<5/4$& $5/4<n<4/3$ & $4/3<n<\frac{1}{2}(1+\sqrt{3})$ & $n>\frac{1}{2}(1+\sqrt{3})$ \\ \mr
 $\mathcal{A}$ & saddle& saddle& saddle& saddle\\
 $\mathcal{B}$   & saddle& repellor& repellor& repellor\\
 $\mathcal{C}$ &repellor&saddle&saddle& attractor\\
 $\mathcal{D}$  &spiral&  spiral & attractor & saddle \\
 $\mathcal{E}$  &saddle& saddle& saddle& saddle\\
 $\mathcal{F}$ & saddle& saddle& saddle& saddle\\
\br
\end{tabular}
\label{tavola punti fissi materia 3a}
\end{table}

 \begin{table}
\caption{Stability of the fixed points  for $R^{n}$-gravity with stiff matter. The term ``spiral" has been
used for  pure attractive  focus-nodes, the term ``anti-spiral" for pure repulsive focus-nodes, and the
term``saddle-focus" for unstable focus-nodes. The point $\mathcal{G}$ has been treated separately because
of the approximations used.}
\centering
\bigskip
\begin{tabular}{cccccc}
\br & $n<\frac{1}{2}(1-\sqrt{3})$ & $\frac{1}{2}(1-\sqrt{3})<n<0$
& $0<n<1/2$& $1/2<n<1$&
 $1<n<\frac{1}{14}(11+\sqrt{37})$  \\ \mr
 $\mathcal{A}$ & saddle& saddle& saddle& saddle&saddle
\\
 $\mathcal{B}$  & saddle& saddle& saddle& saddle& saddle \\
 $\mathcal{C}$ & attractor& saddle& saddle& attractor& repellor  \\
 $\mathcal{D}$ & saddle& attractor &  spiral &  spiral& spiral\\
 $\mathcal{E}$ & saddle& saddle& saddle& saddle& saddle\\
 $\mathcal{F}$ & saddle& saddle& saddle& saddle& saddle\\\mr \mr\\
\end{tabular}

\begin{tabular}{cccccc}
 \br & $\frac{1}{14}(11+\sqrt{37})<n<4/3$ & $4/3<n<\frac{1}{2}(1+\sqrt{3})$  & $\frac{1}{2}(1+\sqrt{3})<n<3/2$& $n>3/2$\\\mr
 $\mathcal{A}$&saddle& saddle& saddle& saddle  \\
 $\mathcal{B}$&saddle& saddle& saddle&saddle \\
 $\mathcal{C}$& saddle&saddle& attractor &attractor \\
 $\mathcal{D}$& spiral&  attractor & saddle& saddle\\
 $\mathcal{E}$&saddle& saddle& saddle& saddle \\
 $\mathcal{F}$&saddle& saddle& saddle& repellor \\
\br
\end{tabular}
\label{tavola punti fissi materia 3c}
\end{table}

\begin{table}
\caption{Stability of the fixed point $\mathcal{G}$. The term ``spiral" has been used for
pure attractive  focus-nodes, the term ``anti-spiral" for pure repulsive focus-nodes, and the
term``saddle-focus" for unstable focus-nodes.}
\centering
\bigskip
\begin{tabular}{ccccccc}
\br & $n\lesssim 0.33$ & $0.33\lesssim n\lesssim 0.35$ &
$0.35\lesssim n\lesssim 0.37$& $0.37\lesssim n\lesssim 0.71$&
  $0.71\lesssim n\lesssim 1$ \\\mr
 $w=0$  &saddle&saddle-focus&saddle-focus&saddle-focus&saddle \\
 $w=1/3$ &saddle&saddle&saddle-focus&saddle-focus&saddle-focus\\
 $w=1$&saddle&saddle&saddle&saddle-focus&saddle-focus\\
\br\\
\end{tabular}
\begin{tabular}{ccccccc}
\br &   $1\lesssim n\lesssim 1.220$ & $1.220\lesssim n\lesssim
1.223$& $1.223\lesssim n\lesssim 1.224$& $1.224\lesssim n\lesssim
1.28$ \\\mr
 $w=0$  &saddle-focus&saddle-focus&saddle-focus &saddle-focus\\
 $w=1/3$ &saddle-focus&saddle-focus&saddle-focus&saddle-focus\\
 $w=1$ &saddle&repellor&saddle&anti-spiral\\
\br\\
\end{tabular}
\begin{tabular}{ccccccc}
\br&   $1.28\lesssim n\lesssim 1.32$& $1.32\lesssim n\lesssim
1.47$& $1.47\lesssim n\lesssim 1.50$& $n\gtrsim 1.50$  \\\mr
 $w=0$ &saddle-focus&saddle& saddle & saddle \\
 $w=1/3$&saddle&saddle& saddle & saddle \\
 $w=1$&anti-spiral&anti-spiral& repellor & saddle \\\br
\end{tabular}
\label{tavola punti fissi materia 4}
\end{table}
The fixed point $\mathcal{A}$ has always a saddle-node character because the eigenvalues of the
linearized matrix do not depend on $n$ and have an alternate sign whatever the value of $w$.

The fixed point $\mathcal{B}$ is an unstable node for $n<1$ and $n>5/4$ if the standard matter is dust
or radiation and a saddle otherwise. In the stiff matter case this point is always a saddle.

The fixed point $\mathcal{C}$  behaves like a pure unstable node for $1<n<5/4$ if the standard matter is
dust or radiation and for $1<n<\frac{1}{14}(11+\sqrt{37})$ in case of stiff matter. A  pure stable character
is instead found for $n<\frac{1}{2}(1-\sqrt{3})$, $1/2<n<1$ and $n>\frac{1}{2}(1+\sqrt{3})$, whatever the value
of the parameter $w$. For all the other values of $n$ this point is a saddle-node.

The fixed point $\mathcal{D}$ has eigenvalues that involve a square root of a function of $n$. This means that when
this function is negative, the eigenvalues are complex and these points behave like focus-nodes. In particular,
for $n<0$ and $n>4/3$ this point is a pure attractive node if $\frac{1}{2}(1-\sqrt{3})<n\leqslant 0$
and $4/3\leqslant n<\frac{1}{2}(1+\sqrt{3})$ or a saddle-node otherwise. For $0<n<4/3$ the eigenvalues
are complex and an analysis of the real parts show that the focus-nodes are always attractive.

The fixed point $\mathcal{E}$ is a saddle-node whatever the values of $n$ and $w$.

The fixed point $\mathcal{F}$ is a saddle-node for every value of $n$ if the matter is dust or radiation.
In the stiff matter case, this fixed point is an unstable node if $n>\frac{3}{2}$.

The calculations of the eigenvalues of the fixed point
$\mathcal{G}$ cannot be performed in an exact way for a general
$n$ because they involve the solution of complete algebraic
equations of order higher than two. Numerical calculations show
that there are complex eigenvalues for $0.33\lesssim n\lesssim
0.71$ and $1 \lesssim n\lesssim 1.33$ if $w=0$, $0.35\lesssim
n\lesssim 1.28$ if $w=1/3$, $0.37\lesssim n\lesssim 1$ and $1.224
\lesssim n\lesssim  1.47$, if $w=1$. For  all these values of $n$,
$\mathcal{G}$ behaves like a saddle focus. For the other values of
$n$ this point is always a saddle-node apart in case of stiff
matter, for which $1.220 \lesssim n\lesssim  1.223$ and $1.47
\lesssim n\lesssim  1.5$. In this case we have a pure repulsive
node \footnote{Also in this case we will not investigate the
behavior of this model in the bifurcations leaving this task for a
future paper.}.

Using the same procedure of the vacuum case, the coordinates of
the fixed points (\ref{punti fissi materia}) can be used to find
the behavior of the scale factor. The equation
(\ref{H-puntifissi}) generalizes to
\begin{equation}\label{H-puntifissi matt}
    \dot{H}=\left(x_{C}+\frac{y_{C}}{n-1}-z_{C}-1\right)H^{2}\;.
\end{equation}
If $n\neq1$ and
 \begin{equation}\label{con matt}
   (n-1)(x_{C}-z_{C}-1)-y_{C}\neq0
\end{equation}
this equation can be integrated to give
\begin{equation}\label{fattore di scala critico materia}
    a=a_{0}(t-t_{0})^{\alpha} \qquad\mbox{with}\qquad \alpha=\left(1-x_{C}-\frac{y_{C}}{n-1}+z_{C}\right)^{-1}\;.
\end{equation}
However, the characterization of the solution in this case
requires also the derivation of the energy density $\rho$ which is
given by the (\ref{rho-dyn var}).

Since for $z_{C}=0$ the solution (\ref{fattore di scala critico
materia}) reduces to the (\ref{a-punti fissi}), the fixed points
$\mathcal{A},\mathcal{B},\mathcal{C},\mathcal{D}$ are associated
with the same solutions given in the vacuum case. The points
$\mathcal{C}$ and $\mathcal{D}$ have $y\neq 0$ and represent
vacuum solution for every value of $n$  and using the cosmological
equations is it possible to show that this is also true for the
solution associated with $\mathcal{A}$ and $\mathcal{B}$.

The equation (\ref{fattore di scala critico materia}) also allows
one to find a solution for $\mathcal{E}$. This point correspond to
a vacuum Milne solution, but since the energy density in this
point is divergent for $n<1$ we consider this point physical only
for $n>1$.

The solution for point $\mathcal{F}$, instead, cannot be found via
the (\ref{fattore di scala critico materia}) because the condition
(\ref{con matt}) is violated. A direct resolution of
(\ref{H-puntifissi matt}) tells us that this point represents a
solution only if $\rho=0$ and $n=3/2$, but this value of the
parameter is forbidden in this case.

The fixed point $\mathcal{G}$ is instead a new
solution with respect to the vacuum case. The scale factor and the
Hubble parameter, are in this case,
\begin{eqnarray}\label{fattore di scala critico G}
&&a=a_{0}\;(t-t_{0})^{\frac{2n}{3(1+w)}}
\end{eqnarray}
and the energy density is given by
\begin{equation}\label{densità punto G}
\rho=\rho_{0} t^{-2n}
\end{equation}
with
\begin{eqnarray}\label{rho0punto G}
\rho_{0}=&&(-1)^{n}\chi(n)3^{-n}2^{2n-1} n^{n} (1+w)^{-2n}\times\\&&(4n-3(1+w))^{n-1}
[2n^{2}(4+3w)-n(13+9w)+3(1+w)]\;.
\end{eqnarray}
Thus  $\mathcal{G}$ represents a power law regime which, for $n>0$, is
an expanding solution with $\rho$ decreasing in time, and for $n<0$ a contracting
solution with $\rho$ increasing in time. If $n$ is close to 1, this solution can be
thought of as an ``almost-Friedman" solution.  Instead, for $n>\frac{3}{2}(1+w)$
this solution corresponds to accelerating expansion.

Unfortunately, since $\rho_{0}$ is a function of $n$ and $w$, $\rho$ is not necessarily positive and
consequently $\mathcal{G}$ is not necessarily a physical point. The values of $n$ for which $\mathcal{G}$ is
physical can be found by solving the inequality $\rho_{0}>0$. We can distinguish two different cases: $n$
belonging to the set of even integers or allowed rational numbers with an even numerator, which we
call $\mathcal{N}_{even}$, and $n$ belonging to the odd integers or allowed rationals with odd
numerator, which we call $\mathcal{N}_{odd}$.
\begin{description}
\item[$\left(n \in \mathcal{N}_{even}\right)$] In this case we require
that $$(4n-3(1+w))[2n^{2}(4+3w)-n(13+9w)+3(1+w)]>0\;.$$
This inequality is satisfied for $0.28\lesssim n\lesssim0.75$ and for $n\gtrsim 1.35$ if $w=0$,
for $0.31\lesssim n\lesssim1$ and $n\gtrsim 1.29$ if $w=1/3$, for $0.35\lesssim n\lesssim 1.22$
and $n\gtrsim 1.5$ if $w=1$. If $n<1$ we should also add the constraint $(4n-3(1+w))\neq 0$ to avoid
divergences. However, the solution of this equation for physical choices of $w$ have values of $n$
which are not forbidden by the above inequalities, or do not correspond to the case $n=1$ .
As a consequence, the matter energy density always remains finite.
\item[$\left(n \in \mathcal{N}_{odd}\right)$]
This case can be split in two different subcases
\begin{description}
\item[$n>0$] for which $$[2n^{2}(4+3w)-n(13+9w)+3(1+w)]<0\;,$$ which leads to
$0.28\lesssim n\lesssim 1.35$ if $w=0$, $0.31\lesssim n\lesssim 1.29$
if $w=1/3$, $0.35\lesssim n\lesssim 1.22$ if $w=1$, the value $n=1$ being excluded.
\item[ $n<0$] for which $$[2n^{2}(4+3w)-n(13+9w)+3(1+w)]>0\;,$$ leads to
$n\lesssim 0$ for $w=0,1/3,1$. \end{description}
\end{description}
It should  be also pointed out that for
$\displaystyle{n=\frac{13+9w\pm\sqrt{73+66w+9w^{2}}}{2(8+6w)}}$,
the density is zero. However these values are not allowed in the
matter case, so the matter density is non-zero for all physical
values of $n$.

Using (\ref{rho-dyn var}), the conditions above can be also read
as conditions on the coordinates of {\cal G}. In particular, the
point {\cal G} will represent a physical point only if
\begin{equation}
\begin{array}{cc}
    z y > 0 & \mbox{for}\qquad \hbox{$n\in {\cal N}_{even}$;} \\
    \left\{%
\begin{array}{ll}
    z < 0, & \hbox{$n > 0$} \\
    z > 0, & \hbox{$n < 0$} \\
\end{array}%
\right.    &  \mbox{for} \qquad \hbox{$n\in {\cal
N}_{odd}$.} \\
\end{array}%
\end{equation}

Again, it is useful to define the deceleration parameter $q$ and
obtain the regions of the phase space in which correspond to
accelerated evolution. We have
\begin{equation}\label{deceleration par}
    q=-\frac{H'}{H^{2}}-1=z-x-\frac{y}{n-1}
\end{equation}
which means, for $q>0$,
\begin{equation}\label{deceleration constr}
    \left\{%
\begin{array}{ll}
    n>1 \\
    (z-x)(n-1)-y>0\\
\end{array}%
\right.
\left\{%
\begin{array}{ll}
    n<1 \\
    (z-x)(n-1)-y<0\\
\end{array}%
\right.
\end{equation}
Thus, chosen $n$, the phase space is divided in two regions which
correspond to accelerating and decelerating evolution. It is easy
to check that the conditions (\ref{deceleration constr}) are never
satisfied for the points ${\mathcal A},{\mathcal B},{\mathcal D}$
and for ${\mathcal E},{\mathcal F}$ (as we would expect from the
solutions given above). For the point ${\mathcal C}$ the
constraint on $n$ are the same as (\ref{acc c cond}) and for the
point ${\mathcal G}$ we have an accelerated evolution only if
$n>\frac{1}{2}(3+3\omega)$ (expansion) and $n<0$ (contraction).

\subsection{Asymptotic analysis}
As already discussed for the vacuum case, we complete the phase space analysis by investigating the
asymptotic regime. The compactification of the phase space is achieved in the same way as the vacuum case,
the difference being that now we now use spherical polar coordinates $(r,\theta,\phi)$ for which
\begin{equation}\label{coordinate sferiche}
    x\rightarrow r\sin(\theta)\cos(\phi),\qquad y\rightarrow r\sin(\theta)\sin(\phi), \qquad z\rightarrow r\cos(\theta)\;,
\end{equation}
where $r\in[0,\infty[$, $\theta\in[0,\pi]$, $\phi\in[0,2\pi]$. As
in the vacuum case, the actual compactification is achieved
transforming the radial coordinate $r$ in ${\cal R}$. In the limit
${\cal R}\rightarrow 1$ the dynamic equations (\ref{sistema
compatto materia}) become
\begin{eqnarray}
\label{rho'matt}  &&\fl\nonumber{\cal R}' \rightarrow
\frac{1}{4(n-1)}\left\{8(n-1)\cos^{3}(\theta)
-2(n-1)\cos(\theta)\sin^{2}(\theta)(\cos(2\theta)-3)-4\cos^{2}(\theta)\sin(\theta)\times\right.\\
\nonumber&&\left. [3(n-1)\cos(\phi)+2\sin(\phi)]+
\sin^{3}(\theta)\left[(9 - 8n)\cos(\phi) - \cos(3\phi) -
7\sin(\phi) \right.\right.\\&&\left.\left.+
\sin(3\phi)\right]\right\}
\end{eqnarray}
\begin{equation}\label{theta'matt}
  \fl{\cal R}\;\theta' \rightarrow
\frac{\cos(\phi)\sin(2\theta)\{2n\sin(\theta)+2\cos(\phi)[(n-1)\cos(\theta)
+\sin(\theta)(\sin(\phi)-\cos(\phi))]\}}{4(n-1)(1-{\cal R})}
\end{equation}
\begin{equation}\label{phi'matt}
\fl {\cal R}\;\phi' \rightarrow \frac{1}{(n-1)(1-{\cal
R})}\{\cos(\phi)\sin(\phi)[(n-1)\cos(\theta)
+\sin(\theta)(\sin(\phi)-\cos(\phi))]\}\;.
 \end{equation}
This system has many interesting features. First of all, the above
equations  are parameterized by $n$ but not by the barotropic
factor $w$. This means that the asymptotic regime is independent
of the nature of the perfect fluid that we choose.  In addition,
since there is no ${\cal R}$ dependence in the (\ref{rho'matt}),
the fixed points are determined only by the angular equations.
Therefore, in order to find the asymptotic fixed points, we can
neglect the ${\cal R}'$ equation and deal only with the
$(\theta',\phi')$ system.

Looking carefully at the system (\ref{theta'matt}-\ref{phi'matt}) we realize that it presents a non-trivial
structure because the polynomials in the R.H.S are not prime to each other. This means that we have an infinite
number of fixed points due to the zeros of the common factor. These points will define ``fixed subspaces" of the
total phase space in which every single point has, in principle, its own stability.

Setting $\theta'=0$ and $\phi'=0$, we obtain the coordinates of
the fixed points. Although we obtain 26 independent solutions,
many of them are associated with the same fixed point on the
sphere due to the periodic nature of the trigonometric functions
in (\ref{theta'matt}-\ref{phi'matt}) and the peculiar topology of
the sphere. The results are summarized in Table \ref{tavola punti
fissi asintotici materia 4}.

\begin{table}
\caption{Coordinates of the asymptotic fixed points  and relative
growing rates in non-vacuum $R^{n}$-gravity. Here ``ind."
represents an indeterminate form.} \centering
\bigskip
\begin{tabular}{cclccc}
\br &&&\centre{3}{\underline{\ \ Growth Rates\ \ }}\\
Point &$(\theta,\phi)$& Coordinates & $y/x$ & $z/x$ & $y/z$  \\\mr
 $\mathcal{A}_{\infty}$ &  $(0,0)$ & $x,y\rightarrow 0\;,z\rightarrow+\infty $&0&$\infty$ &0\\
 $\mathcal{B}_{\infty}$ &  $(\pi,\pi)$&$x,y\rightarrow 0\;,z\rightarrow-\infty $ &0&$\infty$&0\\
 $\mathcal{C}_{\infty}$  &  $(\pi/2,0)$& $y,z\rightarrow 0\;,x\rightarrow+\infty $&0&1&ind.\\
 $\mathcal{D}_{\infty}$ &  $(-\pi/2,\pi)$& $y,z\rightarrow 0\;,x\rightarrow-\infty $&0&1&ind.\\
 $\mathcal{E}_{\infty}$ &  $(\pi/4,0)$&$x,z\rightarrow +\infty\;,y\rightarrow 0 $&0&1&0\\
 $\mathcal{F}_{\infty}$ &  $(3\pi/4,\pi)$&$x,z\rightarrow -\infty\;,y\rightarrow 0 $&0&1&0\\
 $\mathcal{G}_{\infty}$ &   $(\pi/2,\pi/4)$&$x,y\rightarrow +\infty \;,z\rightarrow 0$&1&0&$\infty$\\
 $\mathcal{H}_{\infty}$ & $(\pi/2,5\pi/4)$ &$x,y\rightarrow -\infty \;,z\rightarrow 0$&1&0&$\infty$\\\br
 Line $L1$&  $(\theta,\pi/2)$&$y,z\rightarrow +\infty \;,x\rightarrow 0$&$\infty$&$\infty$&$\tan(\theta)$\\
  Line $L2$ &  $(\theta,3\pi/2)$&$y,z\rightarrow -\infty \;,x\rightarrow 0$&$\infty$&$\infty$&$\tan(\theta)$ \\\br
 $\mathcal{I}^{1}_{\infty}$ &  $( \arctan(n-1),\pi/ 2)$ &$y,z\rightarrow +\infty \;,x\rightarrow 0$
 &$\infty$&$\infty$&$n-1$\\
 $\mathcal{I}^{2}_{\infty}$ &   $(-\arctan(n-1),\pi/2)$&$y,z\rightarrow +\infty \;,x\rightarrow 0$&
 $\infty$&$\infty$&$n-1$\\
 $\mathcal{I}^{3}_{\infty}$ &   $( \arctan(n-1),3\pi/2)$&$y,z\rightarrow -\infty \;,x\rightarrow 0$&
 $\infty$&$\infty$&$n-1$\\
 $\mathcal{I}^{4}_{\infty}$ &  $(-\arctan(n-1),3\pi/2)$&$y,z\rightarrow -\infty \;,x\rightarrow 0$&
 $\infty$&$\infty$&$n-1$\\
\br
\end{tabular}
\label{tavola punti fissi asintotici materia 4}
\end{table}
\begin{table}
\caption{Coordinates, eigenvalues and value of $r'$ of the ordinary asymptotic fixed points in non-vacuum $R^{n}$-gravity.
}
\centering
\bigskip
\begin{tabular}{cccccc}
\br Point & $(\theta,\phi)$& Eigenvalues $\theta-\phi$ & $r'$& Behavior\\
\mr
 $\mathcal{A}_{\infty}$ &  $(0,0)$&$\left[1,1\right]$& 2&$|{\cal{N}}-{\cal N}_{\infty}|=\left[C_{1}\pm\frac{C_{0}}{2}(t-t_{0})\right]^{2}$\\
 $\mathcal{B}_{\infty}$ &  $(\pi,\pi)$&$\left[-1,-1\right]$& -2&$|{\cal{N}}-{\cal N}_{\infty}|=\left[C_{1}\pm\frac{C_{0}}{2}(t-t_{0})\right]^{2}$\\
 $\mathcal{C}_{\infty}$  &  $(\pi/2,0)$& $\left[1,\frac{1}{n-1}\right]$& -2 &$|{\cal{N}}-{\cal N}_{\infty}|=\left[C_{1}\pm\frac{C_{0}}{2}(t-t_{0})\right]^{2}$\\
 $\mathcal{D}_{\infty}$ &  $(-\pi/2,\pi)$& $\left[-1,\frac{1}{1-n}\right]$& 2 &$|{\cal{N}}-{\cal N}_{\infty}|=\left[C_{1}\pm\frac{C_{0}}{2}(t-t_{0})\right]^{2}$\\
 $\mathcal{E}_{\infty}$ &  $(\pi/4,0)$& $\left[-\frac{\sqrt{2}}{2},\frac{n\sqrt{2}}{2(n-1)}\right]
 $&$-\frac{\sqrt{2}}{2}$&$a=a_{0}\exp[\lambda(t-t_{0})]$\\
 $\mathcal{F}_{\infty}$ &  $(3\pi/4,\pi)$ &$\left[\frac{\sqrt{2}}{2},\frac{n \sqrt{2} }{2(1-n)}
 \right]$&$\frac{\sqrt{2}}{2}$&$a=a_{0}\exp[\lambda(t-t_{0})]$\\
 $\mathcal{G}_{\infty}$ &   $(\pi/2,\pi/4)$& $\left[\frac{\sqrt{2} }{2(1-n)},\frac{n\sqrt{2} }
 {2(n-1)}\right]$& $\frac{(1-2n)\sqrt{2} }{2(1-n)}$ &
 $|{\cal N}-{\cal N}_{\infty}|= \left[C_{1}\pm C_{0}\left|\frac{n-1}{2n-1}\right|(t-t_{0})\right]^{\frac{2n-1}{n-1}}$\\
 $\mathcal{H}_{\infty}$ & $(\pi/2,5\pi/4)$ &$\left[\frac{\sqrt{2} }{2(n-1)},\frac{n\sqrt{2} }
 {2(1-n)}\right]$& $\frac{(2n-1)\sqrt{2} }{2(1-n)}$& $|{\cal N}-{\cal N}_{\infty}|= \left[C_{1}\pm C_{0}\left|\frac{n-1}{2n-1}\right|
 (t-t_{0})\right]^{\frac{2n-1}{n-1}}$\\
\br
\end{tabular}
\label{tavola punti fissi asintotici materia 5}
\end{table}

\begin{table}
\caption{Stability of the ordinary asymptotic fixed points  for $R^{n}$-gravity with matter.
These results are independent of the value of $w$.}
\centering
\bigskip
\begin{tabular}{cccc}
 \br Point & $n<0$ & $0<n<1$  & $n>1$\\\mr
 $\mathcal{A}_{\infty}$& repellor& repellor & repellor \\
$\mathcal{B}_{\infty}$& attractor& attractor & attractor \\
 $\mathcal{C}_{\infty}$&saddle& saddle & saddle \\
 $\mathcal{D}_{\infty}$& saddle& saddle& saddle   \\
 $\mathcal{E}_{\infty}$& saddle & attractor & saddle\\
 $\mathcal{F}_{\infty}$ &saddle & repellor & saddle\\
 $\mathcal{G}_{\infty}$&saddle& saddle& saddle\\
$\mathcal{H}_{\infty}$ &saddle& saddle& saddle\\
\br
\end{tabular}
\label{tavola punti fissi asintotici materia 6}
\end{table}
We can distinguish a first set of eight ordinary fixed points
whose coordinates do not depend on the value of $n$ and are
therefore common to all the $R^{n}$ models, two ``fixed lines"
$L1$ and $L2$ again independent of $n$ and other four fixed
points on $L1$ and $L2$ whose coordinates depend on $n$.

The solutions associated with the first four
($\mathcal{A}_{\infty}$-$\mathcal{D}_{\infty}$) can be obtained
using the same kind of procedure used in the vacuum case, we can
obtain their solutions (see Table \ref{tavola punti fissi
asintotici materia 5}).

For the points $\mathcal{E}_{\infty}$ and $\mathcal{F}_{\infty}$,
$y\rightarrow 0$ and $x,z\rightarrow\pm\infty$ at the same rate
($x/z\rightarrow 1$). For these points  equation
(\ref{H-puntifissi matt}) reduces to $\dot{H}=0$ which corresponds
to the de Sitter solution $a=a_{0}\exp[\lambda(t-t_{0})]$ where
$\lambda$ is an integration constant.

For the points $\mathcal{G}_{\infty}, \mathcal{H}_{\infty}$,
$z\rightarrow 0$ and $x,y\rightarrow\pm \infty$ at the same rate
($x/y\rightarrow 1$). In this limit the  first and third equation
of the system (\ref{sistema compatto materia}) reduce to
\begin{eqnarray}\label{asy sys matter}
\nonumber x'\rightarrow-x^2\;,
\end{eqnarray}
which admits the solution
\begin{eqnarray}
    &&x=z\rightarrow\frac{1}{{\cal{N}}-{\cal N}_{\infty}}\;.
\end{eqnarray}
Substituting this result in the (\ref{H-puntifissi matt}) we
obtain
\begin{equation}\label{N asy E, F}
|{\cal N}-{\cal N}_{\infty}|= \left[C_{1}\pm
C_{0}\left|\frac{n-1}{2n-1}\right|(t-t_{0})\right]^{\frac{2n-1}{n-1}}\;,
\end{equation}
as expected from the fact that for $z=0$ the system (\ref{sistema
compatto materia}) reduces to (\ref{eq dyn varridotte}).

The parameter $n$ also determines the stability of the ordinary
fixed points. As we can see from Table \ref{tavola punti fissi
asintotici materia 6}, all but the point $\mathcal{B}_{\infty}$
and $\mathcal{E}_{\infty}$ are unstable for every value of $n$.
The point $\mathcal{B}_{\infty}$ is always an attractor and the
point $\mathcal{E}_{\infty}$ is an attractor only if $0<n<1$.

In addition to the ordinary fixed point, there is a second class
of fixed points defined by two `` fixed lines" $L1$ with
$\phi=\pi/2$ and $L2$ with $\phi=3\pi/2$  in the $(\theta, \phi)$
plane. The solution associated with the fixed lines can  be
obtained with the same method used above. We obtain
\begin{equation}\label{sol linee}
|{\cal{N}}-{\cal
N}_{\infty}|=\left[C_{1}\pm\frac{C_{0}}{2}(t-t_{0})\right]^{2}\;,
\end{equation}
For the points on $L1$ and $L2$, one of the eigenvalues of the
associated linearized matrix is null so that the system is
structurally unstable. This means that the linearization is
inappropriate and the local phase portrait is determined
principally by the non-linear terms. In the presence of structural
instability there is no simple way to devise the stability
properties of a fixed point and we are forced to study directly
small perturbations around the fixed lines. Developing the
(\ref{theta'matt}-\ref{phi'matt}) at first order about the generic
point $(\pi/2,\theta_{0})$, we obtain
\begin{eqnarray}\label{eq perturbazione linea 1}
   &&\delta \theta'=\delta\phi \frac{n \sin^{2}(\theta_{0})\cos(\theta_{0})}{n-1},\\
   &&\delta\phi'= \delta\phi\frac{(n-1)\cos(\theta_{0})-\sin(\theta_{0})}{1-n}\;.
\end{eqnarray}
This system admits the solutions
\begin{eqnarray}\label{soluzioni perturbazione linea 1}
   &&\fl \delta \theta=\pm\frac{A_{1} n \sin^{2}(\theta_{0})\cos(\theta_{0})}{ \sin(\theta_{0})+(1-n)\cos(\theta_{0})}\;\exp\left[\pm\frac{(n-1)\cos(\theta_{0})-\sin(\theta_{0})}{1-n}\;{\cal N}\right]+B_{1},\\
   &&\fl\delta\phi= A_{1}\; \exp\left[\pm\frac{(n-1)\cos(\theta_{0})-\sin(\theta_{0})}{1-n}\;{\cal N}\right],
\end{eqnarray}
where the $``\pm"$ is connected with the sign of the perturbation $\delta\phi$ and $A_{1}$ and $B_{1}$ are integration
constants.  Now, if we look at the value of $r'$ at  the generic point $(\pi/2,\theta_{0})$ we obtain
\begin{equation}\label{r'pi/2}
    r'|_{\phi=\pi/2}= 2\frac{(n-1)\cos(\theta)-\sin(\theta)}{n-1}
\end{equation}
and discover that the stability of the fixed point depends only
on the sign of the term $\displaystyle{\frac{(n-1)\cos(\theta_{0})
+\sin(\theta_{0})}{n-1}}$. If ${\cal N}$ increases with time (i.e.
the model is expanding) and the perturbation $\delta\phi$ is
positive, the points on the fixed line $\phi=\pi/2$ are saddle
nodes for every value of $\theta$ and $n$. Conversely, if the
perturbation $\delta\phi$ is negative, the points on the fixed
line $\phi=\pi/2$ are pure attractors if
$\arctan(n-1)<\theta<\pi/2$ for $n>1$ and
$-\pi/2<\theta<\arctan(n-1)$ for $n<1$ and are pure repellors
otherwise. This behavior is reversed in case of contraction
(${\cal N}$ decreases with $t$).

 Similarly, expanding the equations
(\ref{theta'matt}-\ref{phi'matt}) at first order about the generic
point $(3\pi/2,\theta_{0})$ we obtain
\begin{eqnarray}\label{eq perturbazione linea 2}
   &&\delta \theta'=\delta\phi \frac{n \sin^{2}(\theta_{0})\cos(\theta_{0})}{1-n},\\
   &&\delta\phi'= \delta\phi\frac{\sin(\theta_{0})+(n-1)\cos(\theta_{0})}{1-n}\;.
\end{eqnarray}
This system admits the solutions
\begin{eqnarray}\label{soluzioni perturbazione linea 2}
   &&\fl\delta \theta=\pm\frac{A_{2} n \sin^{2}(\theta_{0})\cos(\theta_{0})}{ \sin(\theta_{0})
   +(n-1)\cos(\theta_{0})}\;\exp\left[\pm\frac{\sin(\theta_{0})+(n-1)\cos(\theta_{0})}{1-n}\;{\cal N}\right]+B_{2},\\
   &&\fl\delta\phi= A_{2}\; \exp\left[\pm\frac{\sin(\theta_{0})+(n-1)\cos(\theta_{0})}{1-n}\;{\cal N}\right],
\end{eqnarray}
where the $``\pm"$ is connected with the sign of the perturbation $\delta\phi$ and $A_{2}$ and $B_{2}$ are
integration constants.  Again, if we look at the value of $r'$ at the generic point $(\pi/2,\theta_{0})$
\begin{equation}\label{r'3pi/2}
    r'|_{\phi=3\pi/2}= 2\frac{(n-1)\cos(\theta)+\sin(\theta)}{n-1},
\end{equation}
we discover that the stability of the fixed point depends only on
the sign of the term $\displaystyle{\frac{(n-1)\cos(\theta_{0})
+\sin(\theta_{0})}{n-1}}$. If ${\cal N}$ increases with time (i.e.
the model is expanding) the perturbation $\delta\phi$ is positive,
the points of the fixed line $\phi=3\pi/2$ are saddle nodes for
every value of $\theta$ and $n$. Conversely, if the perturbation
$\delta\phi$ is negative, the points of the fixed line
$\phi=3\pi/2$ are pure attractors if $-\pi/ 2<\theta<\arctan(1-n)$
for $n>1$ and $\arctan(1-n)<\theta<\pi/2$ for $n<1$ and are pure
repellors otherwise. This behavior is reversed in case of
contraction.

Finally, there are four special points given by
$(\pm\arctan(n-1),\pi/ 2+k\pi)$ with $k=0,1$. These points belong
to the lines $L1$ and $L2$ and are associated with the solutions
(\ref{sol linee}). The difference is that, in calculating their
stability we find that {\em both} the eigenvalues are null. This
situation is more complex than the previous one, and the stability
analysis requires us to perturb the system at the second order.
Unfortunately, the solution of this system turns out to be
achievable only numerically and hence makes it impossible to give
a general stability analysis. For this reason we will limit
ourselves to giving only plots of the solutions for the point
$(\pi/ 2,\arctan(n-1))$ and for a set of specific values of $n$
(see figures 9-14). It is clear that for all the values of $n$ the
perturbation on $\theta$ approaches a constant. Such a behavior is
due to the fact that the fixed lines are themselves invariant
submanifolds and therefore small perturbations on $\theta$ induce
the state to evolve {\em on} the line towards the nearest
attractor.

It is clear that the global behavior of the non-vacuum
$R^{n}$-gravity cosmological models is not trivial. We have three
different types of transient states which represent, Milne
 and ``Friedman-like" solutions. As in the vacuum case,
for values of $n$ far from one the power law solution
$\mathcal{C}$ is an attractor. When $n$ approaches unity, other
attractors i.e. $\mathcal{D}$ come into play and the stability of
$\mathcal{C}$ changes dramatically. Asymptotically two different
sets of attractors are present. The first one is made up of
ordinary points i.e. $\mathcal{B}_{\infty}$ and, for $0<n<1$,
$\mathcal{E}_{\infty}$; the second one by the arcs
$(\phi=\pi/2,0<\theta<\arctan(n-1))$ and $(\phi=3\pi/
2,\arctan(1-n)<\theta<0)$ for $n>1$ and $\arctan(n-1)<\theta<0$
and $0<\theta<\arctan(1-n)$ for $n<1$.

The most interesting scenario is an orbit that approaches first
the $\mathcal{G}$ point and then the $\mathcal{C}$ solution.
Clearly, the values of $n$ should ensure $\mathcal{G}$ is physical
and not too different from the standard Friedmann solution in
order be consistent with observational constraints. At the same
time, $n$ has to be chosen to ensure that $\mathcal{C}$ is
associated with an accelerating solution in order to get a late
acceleration phase. Moreover, since the planes $y=0$ and $z=0$ are
invariant submanifolds and cannot be crossed by any orbit, we have
to make sure that the $y$ coordinate of the points $\mathcal{C}$
and $\mathcal{G}$ have the same sign. Using the Tables given
above, we can select the value of $n$ for which this is realized:
\begin{eqnarray}\label{valori di n ok}
w=0,1/3 && \qquad\left\{\begin{array}{ccc}
n\lesssim 0.37 & \mbox{if}& n \in \mathcal{N}_{even}\\
1.37\lesssim n\lesssim 2 & \mbox{if}& n \in \mathcal{N}_{odd}
\end{array}\right. \\
w=1 &&\qquad\left\{\begin{array}{ccc}
n\lesssim 0.37 & \mbox{if}  & n \in \mathcal{N}_{even}\\
1.5\lesssim n\lesssim 2 & \mbox{if} & n \in \mathcal{N}_{odd}
\end{array}\right.
\end{eqnarray}

\section{Discussion and Conclusions}\label{conclusioni}

In this paper, we have given a detailed analysis of the behavior of
$R^{n}$-gravity cosmological models using the dynamical system
approach. Defining an effective fluid representing the higher
order corrections to standard GR, it is possible to define a system
of first order differential equations which are
equivalent to the cosmological equations. In this way, using
standard dynamical systems theory, we have obtained a
qualitative description of the global and asymptotic behavior of these
cosmologies.  This allows to achieve a comprehensive (but
qualitative) description of the global features of such models.

Our investigation began with an analysis of the the pure gravity
case in which the phase space is two dimensional. We found five
fixed points, two of them (one of which is double) with
coordinates independent of the parameter $n$ and two whose
position on the plane changes with $n$ . These fixed points
correspond to the four exact solutions listed in Table \ref{tavola
punti fissi vuoto 1}. One of them, namely the solution associated
with the point $\mathcal{C}$, is particularly interesting because
it can give rise to accelerated expansion. This solution has
already been found by solving the field equations directly
\cite{revnostra} and was compared with observational data. Our
analysis shows that, for the values of $n$ compatible with
observations, the solution $\mathcal{C}$ is an attractor.

However, the structure of the phase space shows that once we have
set the initial condition, the sign of the Ricci scalar is
preserved. In other words there is no way in which these models
can evolve from a state $R>0$ to a state $R<0$ or vice versa. Such
a transition would be impossible in GR because $R$ is proportional
to the energy density and a change in its sign would imply a
violation of the energy conditions. In $R^{n}$ gravity, and in a
general $f(R)$ gravity, this is not necessarily true because of
the more complicated relation between $R$ and $\rho$. However, the
main consequence of such a feature is that, in the vacuum phase
space, no global attractor can exist and, in particular,
$\mathcal{C}$ is an attractor only for specific initial
conditions.

The introduction of the matter term in the field equations induces
strong constraints on the parameters. The value of $n$ has to be
either a relative or an odd denominator rational number to ensure
self-consistency of the theory. This was expected:
not all forms of non-minimal coupling  are necessarily compatible
with standard matter. Other constraints come from the fact
that the matter energy density is a function of $n$ and is not
positive definite in general.

The analysis of the dynamical system reveals that four of the
seven fixed points found in this case are generalizations of the
fixed point of the vacuum case and present the same stability properties.
This implies that, for example the constraints on $n$ found in
\cite{curvatura obs} hold also in this case. The last three fixed
points are just transients that, providing $w$ lies in the Zel'dovic
interval, are physical only for specific values of $n$. The fixed point
$\mathcal{G}$ is particularly interesting and may be interpreted as
corresponding to an ``almost-Friedmann" transient phase in the
evolution of the cosmological model.

We found values of $n$ for which the universe undergoes a transient
almost-Friedmann behavior, finally evolving into an accelerating phase.
The existence of such a transient would not only explain the recently observed
non constancy of the deceleration parameter \cite{q non const}, but would also
allow, in principle, a period of structure formation. Such a feature is relevant
because it is not easy to explain how large scale structure is formed in accelerating
universes.

In the asymptotic regime $R^{n}$-gravity possesses number of fixed
points which is the same or even bigger than the number of finite
ones. In the vacuum case, we found that these fixed points
represent essentially an exponential evolution. In two cases the
character of the solution (expanding or contracting) depends on
the value of the parameter $n$. For $n$ in the interval suggested
by the observations we have two attractors for $n<1$ and only one
for $n>1$.

In the matter case, the asymptotic structure is very rich. We have
two different sets of critical points. One of them is made by
single points and one is made by two continuous subsets of the
phase space (the ``fixed lines"). For the  set of points
($\mathcal{A}_{\infty}$-$\mathcal{G}_{\infty}$)  the asymptotic
behaviors are of the same  type of the vacuum case apart a pure de
Sitter evolution when $x$ and $z$ diverge at the same rate
($\mathcal{E}_{\infty}$-$\mathcal{F}_{\infty}$). For the fixed
lines the behavior is of the same type of
($\mathcal{A}_{\infty}$-$\mathcal{D}_{\infty}$).

The results presented in this paper have two key features.
Firstly, the application of the dynamical
system approach to higher order gravity cosmology seems to be a
powerful tool, revealing global features of these models.  Secondly,
the results seem to indicate
that the idea of curvature quintessence i.e. to model the
accelerated expansion phenomenon as a high order gravitational
effect, not only seems viable, but is also in good agreement
with our present understanding of the history of the universe.

\end{document}